\RequirePackage[2020-02-02]{latexrelease}

\documentclass[aps,pra, twocolumn, noshowpacs, floatfix]{revtex4}

\usepackage{wrapfig}
\usepackage{amsfonts}
\usepackage{amssymb}
\usepackage{graphicx}
\usepackage{amsmath}
\usepackage{amsmath}
\usepackage[english]{babel}
\usepackage[mathscr]{euscript}
\usepackage{color}
\usepackage{epstopdf}
\usepackage{footnote}
\usepackage{isotope}
\usepackage{ulem}
\usepackage{float}

\usepackage[export]{adjustbox}

\begin{document}

\title{Dissipative Landau-Zener transitions in a three-level bow-tie model: accurate dynamics with the Davydov multi-D$_2$ Ansatz}

\author{Lixing Zhang$^{1}$, Maxim F. Gelin$^{2,1}$, and Yang Zhao$^{1}$\footnote{Electronic address:~\url{YZhao@ntu.edu.sg}}}

\affiliation{$^{1}$\mbox{School of Materials Science and Engineering, Nanyang Technological University, Singapore 639798, Singapore}\\
$^{2}$\mbox{School of Science, Hangzhou Dianzi University, Hangzhou 310018, China}\\
}

\begin{abstract}
We investigate Landau-Zener (LZ) transitions in the three-level bow-tie model (3L-BTM) in a dissipative environment by using the numerically accurate method of multiple Davydov $\mathrm{D}_{2}$ Ans\"atze.
We first consider the 3L-BTM coupled to a single harmonic mode, study evolutions of the transition probabilities for selected values of the model parameters, and interpret the obtained results  with the aid of the energy diagram method. We then explore the 3L-BTM coupled to a boson bath. Our simulations demonstrate that sub-Ohmic, Ohmic and super-Ohmic boson baths have substantially different influences on the 3L-BTM dynamics, which cannot be grasped by the standard phenomenological Markovian single-rate descriptions. We also describe novel bath-induced phenomena which are absent in two-level LZ systems.
\end{abstract}

\date{\today}
\maketitle

\section{introduction}

The Landau-Zener (LZ) transition, which occurs in a two-level system (TLS) with energy spacing tuned by an external field, appears as a diabatic transition when the energy levels draw near. This widely recognized phenomenon has been replicated through experiments across various physical systems, such as Rydberg lithium atoms in a strong electric field \cite{Rydberg}, accelerated optical lattices \cite{Accelerate_lattice}, and atoms in a periodic potential \cite{Periodic_atom}. Moreover, LZ transition variants can be engineered and implemented in many QED devices to fulfill designated functions \cite{LZ_QED1, LZ_QED2, LZ_QED3}.
Taking into account the dissipative environment of the LZ transition, the driven spin-boson model can be borrowed to describe the dynamics of a dissipative LZ model \cite{LZ_HO1, LZ_HO2, LZ_HO3}. By coupling a superconducting qubit to a transmission line, the dissipative LZ transition can be realized in a lab \cite{SQUID,LZ_review}, and by tuning the spin-bath coupling strength, a coherent-to-incoherent transition may emerge.

The original LZ model captures only the crossing of two energy levels. By changing the number of energy levels and the strength of the external driving field, various avoided crossings can emerge at any points in the energy diagram. This leads to countless novel variants of the LZ  model \cite{LZ_review,MLZ1, MLZ2, Nonlinear_3LZ, OldPaper_3LZ}. Experimental realizations of such models usually require systems with large spins or multiple states, such as nitrogen vacancies in diamond \cite{NV1_3LZ, NV2_3LZ}, triple quantum dots \cite{TQD1_3LZ, TQD2_3LZ}, multiple trap Bose-Einstein condensates \cite{TripleWell_3LZ}, and Fe$_8$ molecular nano-magnets \cite{Fe8_3LZ}.

Among myriad variants of the LZ models, the simple yet representative three-level bow-tie model (3L-BTM) is chosen for study here \cite{OldPaper_3LZ}. The name ``bow-tie'' comes from the energy diagram of the model, where three energy levels join at one point in time. As has already been mentioned, any bare LZ-like model has to be considered as a zero-order approximation, because any quantum system under realistic conditions is coupled to the environment which causes consequential relaxation and decoherence processes in the system \cite{Zurek}. Dissipative variants of the 3L-BTM have been studied, but the environment was often approximated by phenomenological decoherence models \cite{Ashhab16}, effective (non-Hermitian) Hamiltonians \cite{Militello19a,Militello19b} and Markovian Lindblad master equations \cite{Militello19c}.
Such oversimplified treatments of the environment may become insufficient, notably taking into account recent progress in engineering or emulating boson baths with arbitrary spectral densities \cite{Home15,Ustinov23,Pollanen23}.

On the other hand, accurate simulations of multilevel dissipative LZ systems  coupled to realistic boson baths are  rare  {\cite{3LZM1}}, owing to substantial computational challenges. Indeed, conventional methods that are based on master equations of the reduced system density matrices necessitate Hilbert-space truncation to manage computational costs. The quasi-adiabatic path integral (QUAPI) method can, in principle, tackle the problem, and it  was applied to the conventional LZ model in Refs.  \cite{Thorwart09,Thorwart15}. However, considering dissipation in the form of time-correlation functions, this method has excessive memory requirements for large spin systems, such as the 3L-BTM explored in this work.
To address these computational issues, in this study we employ the method of the multiple Davydov Ans\"{a}tze. It has been shown that, with the increasing multiplicity (i.e., with the inclusion of a sufficient number of coherent states in the trial state), the method is capable of delivering a numerical accurate solution to the multidimensional time-dependent Schr\"odinger equation \cite{2023-pers}. With its computational cost manageable and its accuracy benchmarked by several numerically ``exact'' computational protocols, the method of multiple Davydov Ans\"{a}tze has been applied to many physical and chemical problems, such as disordered Tavis-Cummings models \cite{D2_TC}, exciton dynamics in transition metal dichalcogenides \cite{D2_TMD}, and photon delocalization in a Rabi dimer model \cite{D2_RD}. In the present work, the method of multiple Davydov Ans\"{a}tze is used to scrutinize  dynamics of LZ transitions in the dissipative 3L-BTM.

The remainder of this paper is arranged as follows. In Sec.~II, we introduce the model, the theoretical framework of the multi-D$_2$ Ansatz, and the observables of concern. In Sec.~III, we present and discuss various obtained results. Sec.~IV is the Conclusion. The convergence tests that demonstrate accuracy of our calculations can be found in Appendix~\ref{Appendix Z}. Additional pertinent technical details are given Appendix~\ref{Appendix A}.

\section{METHODOLOGY}

\subsection{3L-BTM coupled to a single boson  mode}\label{Discr}
The bare Hamiltonian of the 3L-BTM is akin to that of the original LZ model ($\hbar = 1$ from here onwards):
\begin{eqnarray}\label{EQ1}
{\hat H}_{\rm sp} &=& {vt}{S}_{z}+\Delta{S}_{x} =
 \begin{bmatrix}
     vt & ~\Delta & 0\\
     \Delta &~ 0 & \Delta\\
     0 &~\Delta & -vt
 \end{bmatrix}
\end{eqnarray}
Here $v$ is the scanning velocity, i.e., the rate of change of the external field. $\Delta$ is the tunneling strength between the three states. Compare with the Hamiltonian of the LZ model, the Pauli matrices are replaced with the spin-1 operators $S_z$ and $S_x$\cite{SU2}. \\

A single boson  mode that is coupled to the 3L-BTM is then considered. It is described by the quantum harmonic oscillator Hamiltonian
\begin{eqnarray}\label{EQ2}
{\hat H}_{\rm m} &=& \Omega{\hat a}^{\dagger}{\hat a}
\end{eqnarray}
where $\Omega$ is the frequency of the boson mode, ${\hat a}^{\dagger}$ and ${\hat a}$ are the creation and annihilation operators of the mode, respectively. The experimental realization of the above Hamiltonian usually requires a superconducting
quantum interference device (SQUID)\cite{SQUID,LZ_review}.

The SQUID is coupled to the spin system via the mutual inductance. The coupling Hamiltonian can be written as
\begin{eqnarray}\label{EQ3}
{\hat H}_{\rm cpl} &=& \Lambda({\hat a}^{\dagger}+{\hat a})S_x
\end{eqnarray}
where $\Lambda$ specifies the off-diagonal coupling strength which is related to the strength of the mutual inductance.

The addition of the three terms gives the Hamiltonian of the 3L-BTM coupled to a single boson mode:
\begin{eqnarray}\label{EQ4}
{\hat H}_{\rm sgl} &=& {\hat H}_{\rm sp}+{\hat H}_{\rm m}+{\hat H}_{\rm cpl}
\end{eqnarray}

\subsection{3L-BTM coupled to a boson bath}

In reality, due to for example circuit impedance, the spin system undergoes dissipation/dephasing processes. These effects can be described by the coupling of the bare 3L-BTM to a series of harmonic oscillators mimicking a boson bath:
\begin{eqnarray}\label{EQ5}
{\hat H}_{\rm dsp} &=& {\hat H}_{sp} + \sum_k \eta_k({\hat b}_k^{\dagger}+{\hat b}_k)S_x + \sum_k\omega_k{\hat b}_k^{\dagger}{\hat b}_k
\end{eqnarray}
Here $\eta_k$ is the off-diagonal coupling strength, $\omega_k$ is the frequency, and ${\hat b_k}^{\dagger}$,  ${\hat b_k}$ are the creation and annihilation operators of the bath modes.\\

The bath spectral density function can be written as
\begin{equation}\label{EQ6}
J(\omega)=\sum_{k}(\eta_k)^{2}\delta(\omega -\omega _{k})=2\alpha \omega_c^{1-s}\omega^{s}e^{-\omega/\omega_c}
\end{equation}
where $\alpha$ is the system-bath coupling strength, $\omega_c$ is the cut-off frequency, and $s$ is the exponent that characterizes the bath. If $s <$ 1, the bath is sub-Ohmic; if $s$ = 1, the bath is Ohmic; if $s >$ 1, the bath is super-Ohmic. Ohmic-type baths described by Eq. (\ref{EQ6}) are commonly used to model/emulate  cavity QED devices \cite{QED_Spec_Dens_21}.

For practical simulations, $J(\omega)$ has to be discretized. For small $s$, the coupling strengths of bath modes with different frequencies are unevenly distributed, and a linear discretization scheme may not be suitable. In order to address this problem, we adopt a ``density" discretization scheme, similar to the one proposed in Ref.~\cite{discrete_WHB}. The ``density" discretization scheme can be introduced as follows. 
Firstly, the frequency domain $[0, \omega_m]$ is divided into $N$ intervals $[\omega_{k^\prime},  \omega_{k^\prime+1}]$, where $k^\prime = 0, 1, ..., N-1$,  $\omega_{k=N} \equiv \omega_m$ is the maximum frequency considered, and $N$ is the total number of frequency segments. Now we introduce the continuous density function $\rho (\omega)$ of the discrete modes. The integration of $\rho (\omega)$ from 0 to $\omega_m$ must be equal to $N$:
\begin{equation}\label{EQd1}
\int_{0}^{\omega_m}d\omega \rho (\omega) = N
\end{equation}
To relate $J(\omega)$ of Eq.~(\ref{EQ6}) with $\rho (\omega)$, we construct $\rho (\omega)$ in the following form:
\begin{equation}\label{EQd2}
\rho (\omega) = \frac{N}{\int_{0}^{\omega_m}d\omega^\prime J(\omega^\prime)}J(\omega)
\end{equation}
By doing this, $\rho (\omega)$ becomes proportional to $J(\omega)$.

The boundaries of the intervals $[\omega_{k^\prime},  \omega_{k^\prime+1}]$ are chosen to fulfill the requirement
\begin{equation}\label{EQd3}
\int_{\omega_{k^\prime}}^{\omega_{k^\prime+1}}d\omega \rho (\omega) = 1 , \indent k^\prime = 0, 1, ..., N-1
\end{equation}
Then the equivalent frequency and coupling strength for each interval are obtained via the coarse-grained
treatment \cite{coarse}:
\begin{equation}\label{EQd4}
\eta_k = \sqrt{\int_{\omega_{ k^\prime}}^{\omega_{k^\prime+1}}d\omega J (\omega)},\indent  \omega_k = \frac{\int_{\omega_{k^\prime}}^{\omega_{k^\prime+1}}d\omega J (\omega) \omega}{\eta_k^2}
\end{equation}
It is noted this procedure produces equal coupling strengths for all discretized modes, which is given by the expression
\begin{equation}\label{EQd5}
\eta_k = \sqrt{\frac{N}{\int_{\omega_{k^\prime}}^{\omega_{k^\prime+1}}d\omega J(\omega)}}
\end{equation}

\subsection{The multi-D$_2$ Ansatz}

To obtain the system dynamics, the time-dependent Schr\"odinger equation is solved with the multi-D$_2$ Ansatz in the framework of the time-dependent variational principle \cite{2023-pers}. The multi-D$_2$ Ansatz for ${\hat H}_{\rm dsp}$ can be written as
\begin{eqnarray}\label{EQ9}
|{{\rm D}_{2}^M(t)}\rangle& =& \sum_{n=1}^{M}\sum_s ^{+,-,0}A_{ns}|s\rangle \prod_{k}{\mathcal D}_{nk}|{\rm vac}\rangle \nonumber \\
\end{eqnarray}
Here $M$ is the Ansatz multiplicity, $|{\rm vac}\rangle$ is the vacuum state and ${\mathcal D}_{nk}$ is the displacement operator of the $k$th bath mode, which can be written as
\begin{eqnarray}
{\mathcal D}_{nk} = \exp[\alpha_{nk}b_k^{\dagger}- \alpha_{nk}^{\ast}b_k]
\end{eqnarray}
where $\alpha_{nk}$ is the displacement of the effective bath mode and asterisk denotes complex conjugation. The bath part of the wave function is represented by $M$ coherent states in the Ansatz (i.e., $n$ goes from $1$ to $M$). $|s\rangle$ ($s = +,-,0$) denote  the three states of the 3L-BTM, each of which is assigned with an amplitude $A_{ns}$. $A_{ns}$ and $\alpha_{nk}$ are called the variational parameters. These parameters can be determined through the Euler-Lagrange equation under the Dirac-Frenkel time-dependent variational principle:
\begin{align}\label{EQ10}
\frac{d}{dt}\frac{\partial L}{\partial{\dot{u}}_n^\ast}-\frac{\partial L}{\partial u_n^\ast}=0,~ {{u}}_n \in [A_{ns}, \alpha_{nk}]
\end{align}
with
\begin{eqnarray}\label{EQ11}
L&=&\frac{i}{2}\left[\langle {\rm D}_{2}^{M}(t)|\frac{\overrightarrow{\partial}}{\partial t}|{\rm D}_{2}^{M}(t)\rangle
-\langle {\rm D}_{2}^{M}(t)|\frac{\overleftarrow{\partial}}{\partial t}|{\rm D}_{2}^{M}(t)\rangle\right]\nonumber\\&&-\langle{\rm D}_{2}^{M}(t)|\hat{H}_{\theta}|{\rm D}_{2}^{M}(t)\rangle.
\end{eqnarray}
The collection of the Euler-Lagrange equations for all variational parameters yields the equations of motion (EOMs). The EOMs are essentially first-order differential equations, which can be solved simultaneously via, e.g.,  the 4$^{th}$ order Runge-Kutta method. The complete set of the EOMs is presented in Appendix \ref{Appendix A}.

\subsection{Observables}
According to Eq. (\ref{EQ9}), the  multi-D$_2$ wave function is normalized,
\begin{eqnarray}\label{EQ12}
\langle D_2^M(t) | D_2^M(t) \rangle = \sum_{m,n}^{M}\sum_s^{+, -, 0} A_{ms}^{\ast}(t){A}_{ns}(t)S_{mn}=1
\end{eqnarray}
Here
\begin{eqnarray}\label{EQ13}
S_{mn} &= &\langle{0}|\sum_k{\mathcal D}_{mk}^{\dagger}   {\mathcal D}_{nk}|0\rangle \nonumber \\
&=& {\rm exp}\left[\sum_k\alpha_{mk}^{\ast}\alpha_{nk}-\frac{1}{2}(\lvert{\alpha_{mk}}\lvert^2+\lvert{\alpha_{nk}}\lvert^2)\right]\nonumber \\
\end{eqnarray}
is the Debye-Waller factor.
Expectation value of any operator $Q$ can be evaluated in the multi-D$_2$ framework as
\begin{eqnarray}\label{EQ14}
&&\langle Q(t)\rangle = \langle{{\rm D}_{2}^{M}(t)}|Q|{{\rm D}_{2}^{M}(t)}\rangle
\end{eqnarray}
The LZ transition probabilities are therefore defined as
\begin{eqnarray}\label{EQ15}
{\mathcal P}_{s}(t) = \langle{D_2^M}|{\mathcal P}_{s}|{D_2^M}\rangle = \sum_{m, n}^{M}A_{ms}^{\ast}{A}_{ns}S_{mn}\nonumber \\
\end{eqnarray}
($s = +,-,0$), where ${\mathcal P}_{s} = |s\rangle\langle{s}|$  is the projection operator.  Due to the unitarity of the Hamiltonian dynamics, the sum of the three transition probabilities equals to 1 at any time $t$. If two of the transition probabilities are available, the third one is known automatically.

\section{RESULTS AND DISCUSSION}\label{RD}

In the absence of the coupling to harmonic oscillators, transitions in multilevel LZ models have been extensively studied, and analytical solutions for the asymptotic transition probabilities ${\mathcal P}_{s}(\infty)$ for the 3L-BTM and other multilevel models have been found \cite{OldPaper_3LZ,3LZ_Analytical,dibatic_basis}. If the 3L-BTM  is initialized in $|0\rangle$, then ${\mathcal P}_{+}(t)={\mathcal P}_{-}(t)$ due to the SU2 symmetry of the Hamiltonian (i.e., the tunneling from $|0\rangle$ to $|\pm \rangle$ states is equally likely).

For LZ systems coupled to harmonic oscillators,  the off-diagonal ($S_x$) coupling is qualitatively similar to  tunneling and  the so-induced LZ transitions reveal the oscillator frequency $\Omega$ \cite{LZ_HO1, LZ_HO2}. The diagonal ($S_z$) coupling may affect the transition probability at long times and finite temperatures \cite{finite_T}, but its effect on the system dynamics is almost  trivial at zero temperature. As temperature effects are not considered in this work, we focus on the off-diagonal coupling.

In Sec.~\ref{Om}, we study a simple case of the 3L-BTM  coupled to one harmonic oscillator, and analyze the dynamics and infinite-time transition probabilities. The dissipative regime is explored in Sec.~\ref{Diss}. Computational details and convergence tests of our  multi-D$_2$ calculations can be found in Appendix \ref{Appendix Z}.

Note that the actual values of time and other relevant parameters are decided by the choice of the unit frequency $\omega$, which makes all these parameters dimensionless. The value of $\omega$ has no effect on the dynamic observables.

\subsection{Dynamics of the 3L-BTM coupled a single harmonic mode}\label{Om}

\begin{figure}[ht]
\centerline{\includegraphics[width=90mm]{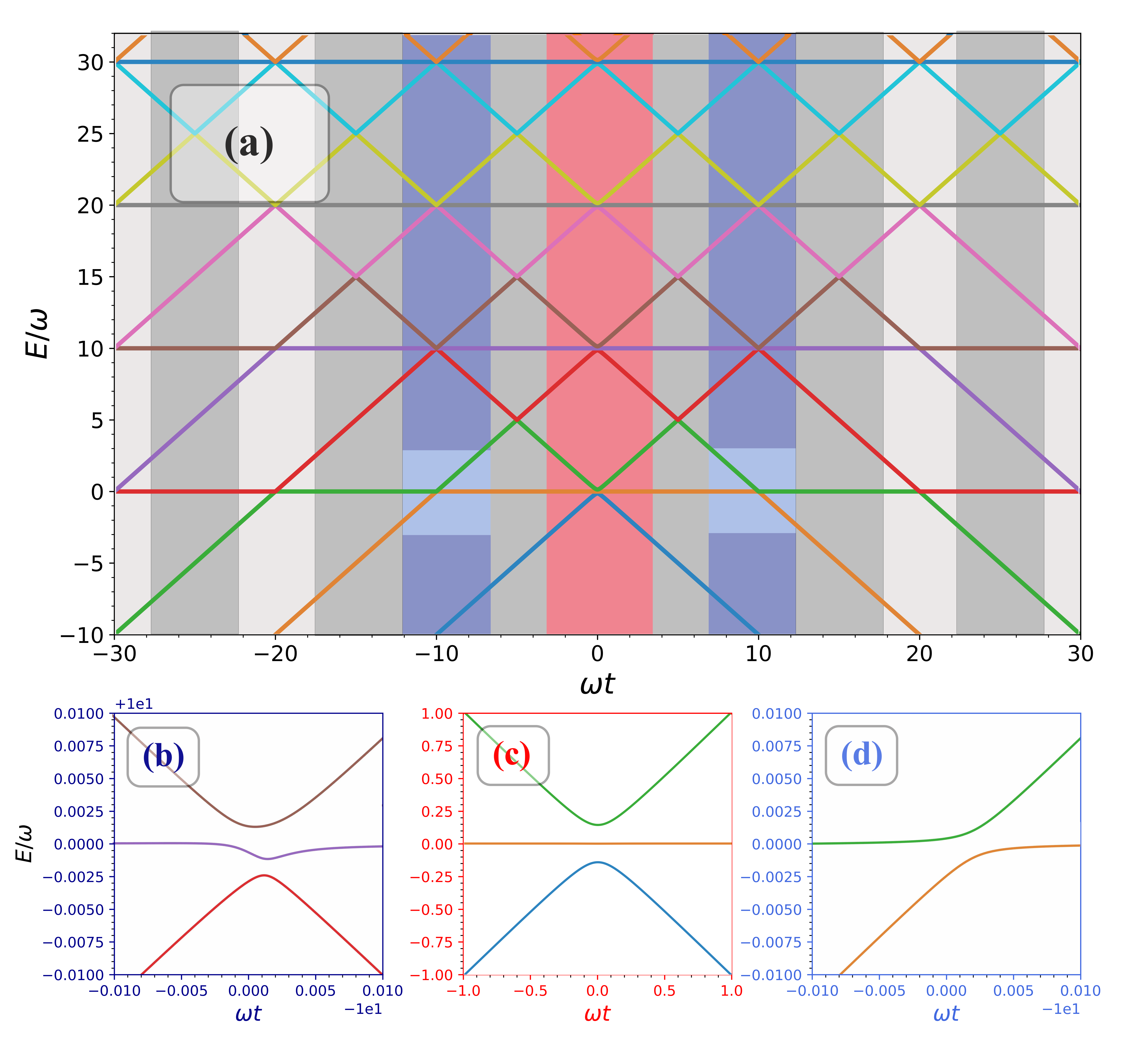}}
\caption{(a): The time evolution of eigenvalues of the  Hamiltonian ${\hat H}_{\rm sgl}$ from $ \omega t = - 30$ to $ \omega t =  30$ for  $\Lambda/\omega = 0.1$, $v/\omega^2 = 1$, $\Delta/\omega = 0.1$, $\Omega/\omega = 10$. The background colors represent different types of the crossings. Gray: forbidden type f1; dark gray: forbidden type f2; dark blue: anti-crossing type a1; light blue: anti-crossing type a2; red: anti-crossing type a3. (b)-(d): Zoom-in plots of three different types of anti-crossings. The frame colors correspond to the background colors in the top panel.}
\label{fig1}
\end{figure}

First, we study the 3L-BTM coupled to a single harmonic oscillator which is described by the Hamiltonian ${\hat H}_{\rm sgl}$ of Eq. (\ref{EQ4}). In Fig.~\ref{fig1}(a), we present the time-dependent energy diagram for this system. The instantaneous (time-dependent) eigenvalues of ${\hat H}_{\rm sgl}$ are plotted from $ \omega t = - 30$ to $\omega t = 30$ for $\Lambda/\omega = 0.1$, $v/\omega^2 = 1$, $\Delta/\omega = 0.1$ and $\Omega/\omega = 10$. The energy diagram is symmetric with respect to $\omega t = 0$. The energy levels can be separated into three groups based on their time gradients $+v, -v$, and $0$, which correspond to the spin states $|{+}\rangle$, $|{-}\rangle$ and $|0\rangle$, respectively. Each group contains numerous parallel energy levels that are separated by $\Omega/\omega$ and correspond to different boson numbers. Only the first few levels are included in Fig.~\ref{fig1}(a).   To distinguish different types of level crossings in Fig.~\ref{fig1}(a),  different background colors are used for each type of crossings.

For crossings with gray and dark gray background colors, named as type f1 and f2 crossings, respectively, LZ transitions are forbidden. Indeed, type f1 crossings include three energy levels: $|{+, m+2n}\rangle,  |{0, m+n}\rangle$ and $|{-, m}\rangle (m \in \mathbb {N}, n \ge 2)$. LZ transitions between these states require emission/absorption of $n$ bosons, which is forbidden since ${\hat H}_{\rm cpl}$ supports $\pm 1$ changes in the boson number only. For type f2 crossings, two energy levels with spin $+$ and $-$ are involved only. Direct LZ transition between these levels are also forbidden, because the $S_x$-operator supports tunneling only between the adjacent spin states.

For crossings with dark blue, light blue and red background colors, named as type a1, a2 and a3 crossings, respectively, LZ transitions are allowed. For type a1 and a2 crossings, LZ transitions induce $\pm 1$ changes in the boson number. The gap opened at these avoided crossings is determined by the coupling strength $\Lambda$. It is noted that the rotating wave approximation (RWA) is not applied in ${\hat H}_{\rm cpl}$. As vibronic levels are bounded from below by the vacuum state, differences exist between type a1 crossings (which involve three energy levels) and type a2 crossings (which involve only two levels). Hence type a1 crossings are dynamically  crucial if  the 3L-BTM is initialized in higher vibronic states, while type a2 crossings are important if the 3L-BTM is initialized in the vacuum state or lower vibronic states. For type a3 crossings, LZ transitions involve a change of spin only. Such spin flips are caused by the tunneling, and the gap opened at the avoided crossing is solely decided by $\Delta$. Type a3 is the only crossing in the bare 3L-BTM. Transitions between $|{+}\rangle$ and $|{-}\rangle$ are forbidden for type f2 crossings, but are allowed for type a3 crossings via the intermediate state $|0\rangle$.

\begin{figure*}[ht]
\centerline{\includegraphics[width=200mm]{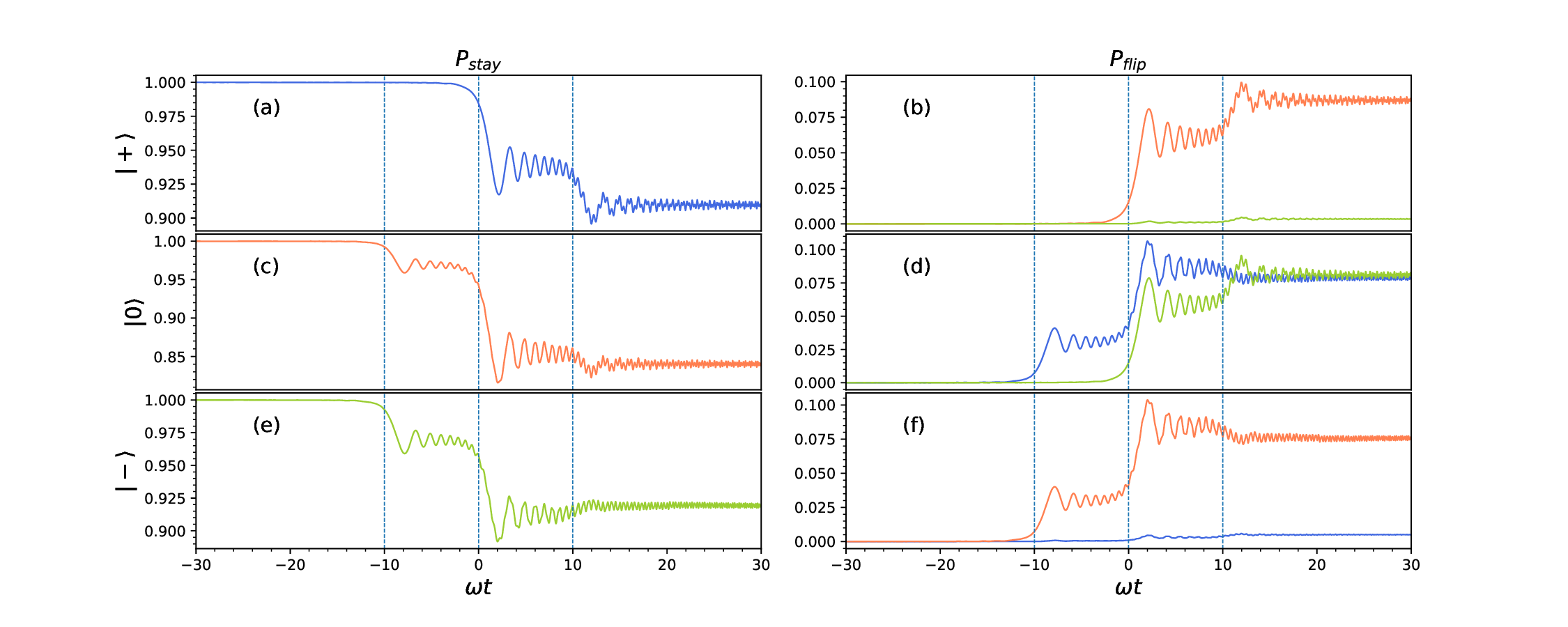}}
\caption{Time evolutions of transition probabilities ${\mathcal P}_{s}(t)$ from $\omega t = -30$ to $\omega t = 30$. The colors distinguish transition probabilities in different states: blue, ${\mathcal P}_{+}(t)$; orange, ${\mathcal P}_{0}(t)$, yellow-green, ${\mathcal P}_{-}(t)$. The left column shows the   probabilities to stay in the originally populated state, while the right column shows the probability of the spin  flip. The rows (from top to bottom) correspond to the initializations of the system in the states $|{+}\rangle$,  $|0\rangle$, $|{-}\rangle$, respectively. All transition probabilities are evaluated for $\Lambda/\omega = 0.1$, $v/\omega^2 = 1$, $\Delta/\omega = 0.1$ and $\Omega/\omega = 10$. }
\label{fig2}
\end{figure*}

Displayed in Fig.~\ref{fig1}(b)-(d) are zoom-in plots of the three types of avoided crossings: a1, a3 and a2.  The frame colors of these plots correspond to the background colors of the crossings in panel Fig.~\ref{fig1}(a).
Fig.~\ref{fig1}(c) shows
type a3 crossing corresponding to $E/\omega = 0 , \omega t = -10$ in the energy diagram. Two gaps of equal size are opened simultaneously in time: between $|{+}\rangle$ and $|0\rangle$, and between $|{-}\rangle$ and $|0\rangle$.  As the result, if the wave function is initialized in $|0\rangle$, the transition probabilities to $|\pm \rangle$ are equal at all times.
Fig.~\ref{fig1}(b) illustrates type a1 crossing  corresponding to $E/\omega = 10, \omega t = -10$ in the energy diagram. Unlike type a3 crossings, the upper and lower gaps neither occur simultaneously in time nor are equal in size, being asymmetric with respect to $\omega t=0$. This breaks the symmetry between the LZ transitions from $|0\rangle$ to $|\pm \rangle$.   Fig.~\ref{fig1}(d), displays type a2 crossing for $E/\omega = 0, \omega t = -10$ ($\omega t = 10$), which involves only a pair of states $|{+}\rangle$ and $|0\rangle$ (or $|{-}\rangle$ and $|0\rangle$). In this case,  transitions in the 3L-BTM are identical to those in the conventional two-level LZ model.

In Fig.~\ref{fig2}, transition probabilities from $\omega t = -30$ to $\omega t = 30$ are presented for different  initializations. The top, middle, and bottom rows correspond to the wave function initialized in $|{+,{\rm vac}}\rangle$, $|{0,{\rm vac}}\rangle$ and $|{-,{\rm vac}}\rangle$, respectively. The left column shows the probability to retain the initial spin direction, $P_{\rm stay}$, and the right column shows the transition probability of the spin flip, $P_{\rm flip}$. The transition probabilities for  different states are designated in different colors, blue: ${\mathcal P}_{+}(t)$, orange: ${\mathcal P}_{0}(t)$ and yellow-green: ${\mathcal P}_{-}(t)$.
The parameters adopted for the plots in Fig.~\ref{fig2} are same as for Fig.~\ref{fig1}: $\Lambda/\omega = 0.1$, $v = 1$, $\Delta/\omega = 0.1$ and $\Omega/\omega = 10$.

In Figs.~\ref{fig2}(a) and (b), the wave function is initialized in $|{+,{\rm vac}}\rangle$. The time evolution of ${\mathcal P}_{\pm}(t)$  can be divided into three stages. The first LZ transition corresponds to type a3 crossing at $\omega t = 0, E/\omega = 0$ in the energy diagram. Clearly,  ${\mathcal P}_{+}(t)$ and ${\mathcal P}_{0}(t)$ change considerably in time, while ${\mathcal P}_{-}(t)$ remains almost zero.
This indicates that the direct transition from $|{+}\rangle$ to $|0\rangle$ is much more probable than the indirect transition from $|{+}\rangle$ to $|{-}\rangle$ via $|0\rangle$. As type a3 crossing does not involve boson states, the boson degrees of freedom remain in the vacuum state $|{{\rm vac}}\rangle$, in agreement with Ref. \cite{bath2}.
At $\omega t = 10$, the wave function goes, simulataneously  through type a2 crossing ($E/\omega = 0$  in the energy diagram), and type a1 crossing  ($E/\omega = 10$  in the energy diagram).
As both crossings are caused by the coupling to the boson mode, the result is decided by the value of this coupling. Since $\Lambda/\omega = 0.1$ is relatively small, indirect transitions from $|{+}\rangle$ to $|{-}\rangle$ are suppressed, and the direct transition from $|{+}\rangle$ to $|0\rangle$ is  dominant. This is a clear indication of the similarity  of the off-diagonal coupling and tunneling in LZ transitions.

\begin{figure*}[ht]
\centerline{\includegraphics[width=220mm]{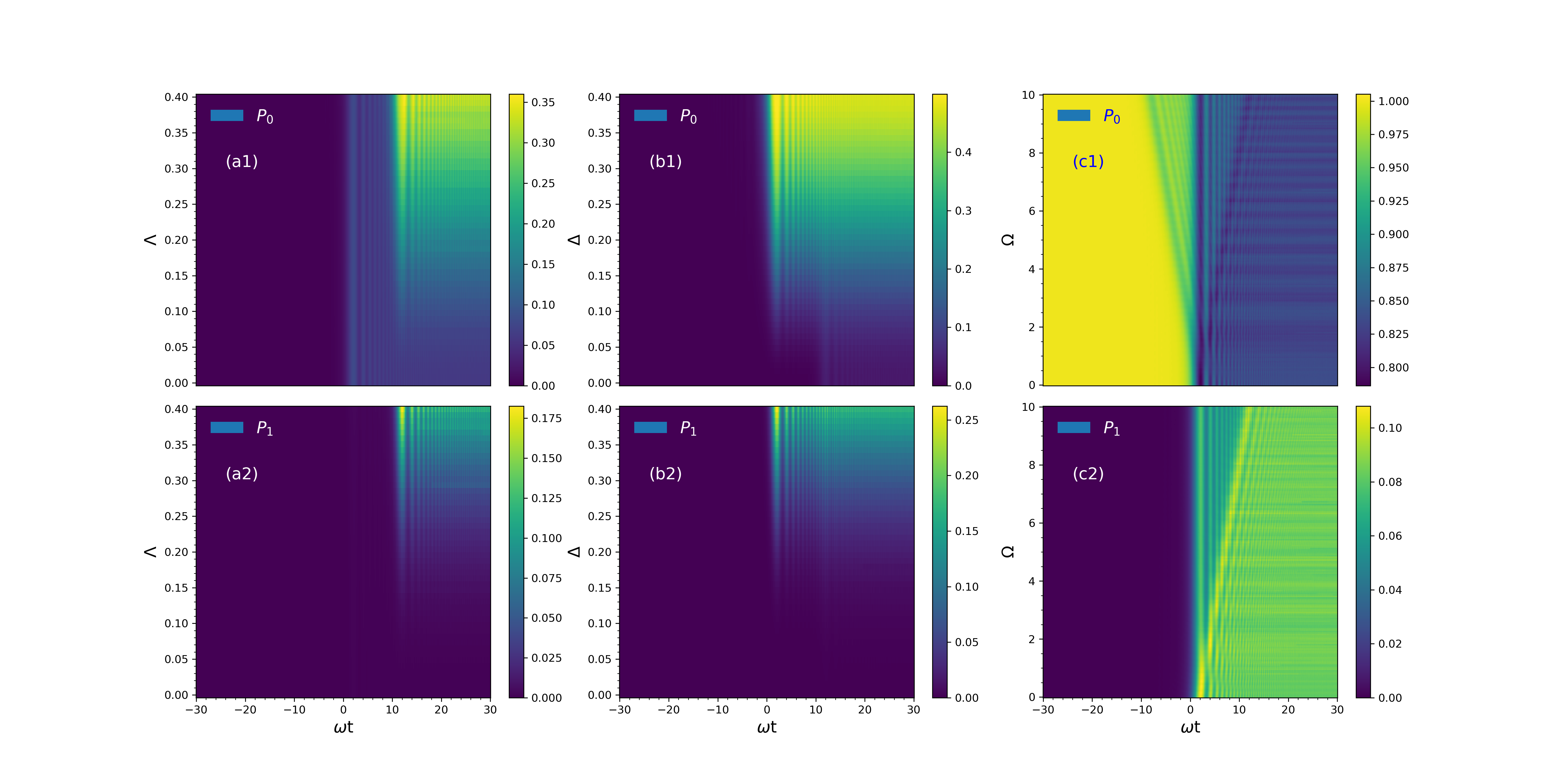}}
\caption{Color maps of time evolutions of LZ transition probabilities  with respect to different system parameters.(a) and (d):  $\Lambda/\omega$ changes from 0 to 0.4, $\Delta/\omega$ = 0.1, $\Omega/\omega$ = 10, the system is initialized in $|{+, {\rm vac}}\rangle$; (b) and (e):  $\Delta/\omega$ changes from 0 to 0.4, $\Lambda/\omega$ = 0.1, $\Omega/\omega$ = 10, the system is initialized in $|{+, {\rm vac}}\rangle$. (c) and (f):  $\Omega/\omega$ changes from 0 to 10, $\Lambda/\omega$ = 0.1, $\Delta/\omega$ = 0.1, the system is initialized in $|{0, {\rm vac}}\rangle$. The upper panels (a), (b), (c) correspond to ${\mathcal P}_{0}(t)$, the lower panels (d), (e), (f) correspond to ${\mathcal P}_{-}(t)$. The scanning velocity is fixed at $v/\omega^2 = 1$.}
\label{fig3}
\end{figure*}

In Figs.~\ref{fig2}(c) and (d), the wave function is initialized in $|{0,{\rm vac}}\rangle$, and the LZ transition at $\omega t = -10$ is of type a2 ($E/\omega = 0$ in the energy diagram). It involves only two levels, i.e., $|{0, {\rm vac}}\rangle$ and $|{+, 1}\rangle$. Hence ${\mathcal P}_{-}(t)$ remains the same before and after this transition, in close similarity with the two-level LZ model~\cite{LZ_HO1}.
The  transition at $\omega t = 0$ has the same origin as the first transition which occurs when the wave function is initialized in $|{+,{\rm vac}}\rangle$.
As the initial state is now $|{0,{\rm vac}}\rangle$ transitions from $|0\rangle$ to $|\pm \rangle$ are equally likely, and increase of the values of ${\mathcal P}_{-}(t)$ and ${\mathcal P}_{+}(t)$ after the transition are the same. The third transition at $\omega t = 10$ is the result of the combination of a single type a2 crossing and multiple type a1 crossings. After this transition, ${\mathcal P}_{-}(t)$ and ${\mathcal P}_{+}(t)$ converge asymptotically to almost the same values. This is due to the fact that
the energy diagram in Fig.~\ref{fig1} is symmetric relative to $\omega t = 0$.
Since the wave function is initialized in $|{0,{\rm vac}}\rangle$, the ensuing 3L-BTM dynamics can be understood as a superposition of the dynamics of a pair of  two-level systems (see Figs.~\ref{fig2}(c) and (d)).

In Figs.~\ref{fig2} (e) and (f), the wave function is initialized in $|{-,{\rm vac}}\rangle$. Similar to the  initialization in $|{+,{\rm vac}}\rangle$, indirect transitions from $|{-}\rangle$ to  $|{+}\rangle$ are suppressed. However, the system initialized in $|{-,{\rm vac}}\rangle$ encounters type a1 crossing at $\omega t = -10$ ($E/\omega = 10$ in the energy diagram). As type a1 crossing involves three energy levels, the first LZ transition at $\omega t = -10$ affects populations of all three states.
As the transition from $|{-}\rangle$ to  $|{+}\rangle$ is suppressed, only a trivial change in ${\mathcal P}_{+}(t)$ is seen after $\omega t = -10$. This is at variance with the outcome of the first LZ transition in the case of $|{0,{\rm vac}}\rangle$ initialization, where only two energy levels are involved. The second transition at $\omega t = -10$ is  similar to the transition which occurs after the wave function is initialized in $|{+,{\rm vac}}\rangle$ or $|{0,{\rm vac}}\rangle$.
For the third LZ transition, the situation is more involved. If the wave function is initialized in $|{0,{\rm vac}}\rangle$, the asymmetry of type a1 crossing is cancelled due to the time symmetry of the energy diagram. If the wave function is initialized in $|{-,{\rm vac}}\rangle$ or $|{+,{\rm vac}}\rangle$, this does not happen. For example, the LZ transition at $\omega t = 10$ increases ${\mathcal P}_{+}(t)$ and ${\mathcal P}_{-}(t)$ (${\mathcal P}_{0}(t)$ and ${\mathcal P}_{-}(t)$) and  decreases ${\mathcal P}_{0}(t)$ (${\mathcal P}_{+}(t)$) if the wave function is initialized in  $|{-,{\rm vac}}\rangle$ ($|{+,{\rm vac}}\rangle$).

\begin{figure*}[ht]
\centerline{\includegraphics[width=220mm]{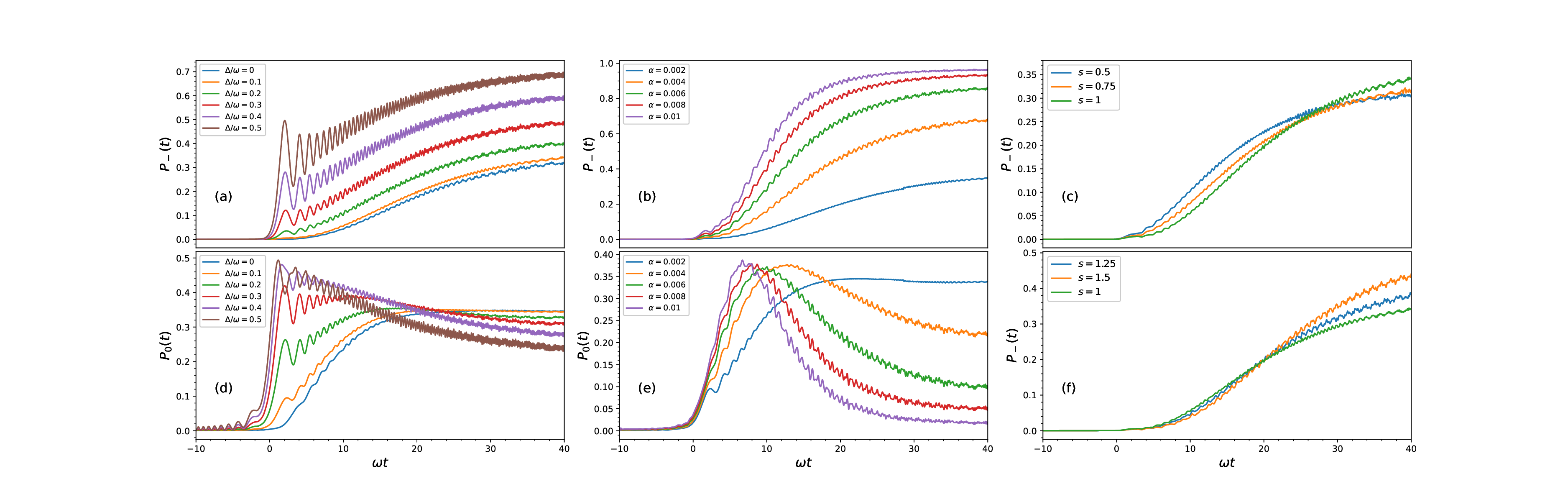}}
\caption{Time evolutions of the transition probabilities ${\mathcal P}_{-}(t)$ (upper pannels) and ${\mathcal P}_{0}(t)$ (lower panels) from $\omega t = -10$ to $\omega t = 40$ for the 3L-BTM coupled to a harmonic bath with the spectral density of Eq. (\ref{EQ6}). For (a) and (d), $\alpha = 0.002$, $s = 1$, and $\Delta/\omega$ is changed from 0 to 0.5 with a step of 0.1. For (b) and (e), $\Delta/\omega = 0.1$, $s = 1$, and $\alpha$ varies from 0.002 to 0.01 with a step of 0.002. For (a) and (b), the transition probability to $|{-}\rangle$ is plotted, whereas for (d) and (e), the transition probability to $|0\rangle$ is plotted. For (c) and (f), $\alpha = 0.002$ and $\Delta/\omega = 0.1$, and the transition probability to $|{-}\rangle$ is plotted. For (c), $s = 0.5, 0.75, 1$ and the bath is sub-Ohmic. For (f), $s = 0.5, 0.75, 1$and the bath is super-Ohmic. For all figures, $\omega_c/\omega = 10$.}
\label{fig4}
\end{figure*}

To grasp the 3L-BTM dynamics in different parameter regimes, Fig.~\ref{fig3} displays time dependent LZ transition probabilities for different off-diagonal coupling strengths $\Lambda/\omega$, tunneling strengths $\Delta/\omega$, and boson mode frequencies $\Omega/\omega$. For the left and middle columns, the wave function is initialized in $|{+, {\rm vac}}\rangle$, and for the right column it is initialized in $|{0, {\rm vac}}\rangle$. The scanning velocity is fixed at $v/\omega^2 = 1$.

In Figs.~\ref{fig3} (a) and (d), $\Lambda/\omega$ is changed from 0 to 0.4, $\Omega/\omega=10$, and $\Delta/\omega=0.1$. The contour plots in Figs.~\ref{fig3} (a) and (d) are clearly separated by two vertical lines at $\omega t = 0$ and $\omega t = 10$ which correspond to the same LZ transitions which appear in Fig.~\ref{fig2} (b). As has been mentioned, the first LZ transition is governed by the tunneling $\Delta/\omega$, while the second LZ transition is governed by the coupling  $\Lambda/\omega$ to the boson mode.
As $\Delta/\omega$ is fixed,  the impact of the first LZ transition  is uniform for all transition probabilities, while the influence of the
second LZ transition at $\omega t = 10$ depends significantly on $\Lambda/\omega$.
There are threshold values of $\Lambda/\omega$ below which ${\mathcal P}_{0}(t)$ and ${\mathcal P}_{-}(t)$ do not change substantially during the second LZ transition.
These threshold values are around 0.05 (0.11)  for ${\mathcal P}_{0}(t)$ (${\mathcal P}_{-}(t)$). Between the two threshold values, that is for $0.05< \Lambda/\omega< 0.11$, the LZ transition changes ${\mathcal P}_{0}(t)$, but does not change ${\mathcal P}_{-}(t)$.
This parameter regime corresponds to the situation illustrated by Figs.~\ref{fig2}(a) and (b), where the direct transition from $|{+}\rangle$ to $|0\rangle$ is much stronger than  the indirect transition from $|{+}\rangle$ to $|{-}\rangle$. If $\Lambda/\omega> 0.11$, the second LZ transition  significantly enhances ${\mathcal P}_{-}(t)$ and ${\mathcal P}_{0}(t)$.

Similar phenomenon can be seen in Figs.~\ref{fig3}(b) and (e), where $\Delta/\omega$ is changed from 0 to 0.4,  $\Omega/\omega=10$, and $\Lambda/\omega=0.1$.
There are also threshold values of
$\Delta/\omega=0.02$ (0.13) below which ${\mathcal P}_{0}(t)$ (${\mathcal P}_{-}(t)$) do not change substantially after the LZ transition at $\omega t = 0$. If
$\Delta/\omega> 0.13$, the LZ transition at $\omega t = 0$ substantially increases ${\mathcal P}_{-}(t)$ and ${\mathcal P}_{0}(t)$.
For large $\Delta/\omega$, for example for $\Delta/\omega = 1$, indirect transitions from $|{+}\rangle$ to $|{-}\rangle$ dominate over direct transitions from $|{+}\rangle$ to $|0\rangle$. Hence the population transfer occurs mainly between the  $|{+}\rangle$ and $|{-}\rangle$ states and  ${\mathcal P}_{0}(t)$ remains almost unchanged. This regime takes place in the 3L-BTM without boson coupling for large  $\Delta/\omega$ \cite{large_Delta}.

In Figs.~\ref{fig3} (c) and (f), $\Omega/\omega$ varies from 0 to 10, while $\Lambda/\omega=\Delta/\omega=0.1$. The wave function is now initialized in $|{0, {\rm vac}}\rangle$. Clearly, the boson mode causes additional   LZ transitions which occur at $\omega t = -\Omega/\omega$ in ${\mathcal P}_{0}(t)$ and at $\omega t = \Omega/\omega$ in ${\mathcal P}_{-}(t)$.

Interestingly, changes in $\Omega/\omega$ cause periodic variations in the steady state transition probabilities and -- for sufficiently large $\Omega/\omega$ --  periods and amplitudes of these variations substantially decrease and become negligible.
These variations are caused by the interference of LZ transitions at $\omega t = \pm \Omega/\omega$ and $\omega t = 0$.
For definiteness, let us consider ${\mathcal P}_{0}(t)$. If the time separation between the two LZ transitions is small, the 3L-BMT has no time to evolve after the first transition at $\omega t = -\Omega/\omega$, and  the second transition at $\omega t = 0$ kicks in shortly after the first transition. Depending on the temporal separation between the two transitions, the first transition quenches at different times and yields different steady-state transition probabilities. Consequently, the steady-state transition probabilities vary with  $\Omega/\omega$.
If  the temporal separation between the two LZ transition is large enough, the system after the first transition at $\omega t = -\Omega/\omega$ will be fully relaxed before the second transition occurs at $\omega t = 0$. This eliminates oscillations in the steady-state transition probabilities with respect to different $\Omega/\omega$. However, as the emergence of this phenomenon hinges upon the LZ transition occurring at $\omega t = -\Omega/\omega$, it does not occur if the wave function is initiated in $|{+, {\rm vac}}\rangle$.

\subsection{Dynamics of the 3L-BTM coupled to a dissipative bath}\label{Diss}

In the previous section, we considered the 3L-BTM  coupled to a single harmonic mode. Here the 3L-BTM is coupled to an Ohmic-type bath described by the Hamiltonian ${\hat H}_{\rm dsp}$ of Eq.~(\ref{EQ5}).
In Figs.~\ref{fig4}(a), (b), (d), and (e), effects of $\Delta/\omega$ and $\alpha$ on the dynamics of the 3L-BTM coupled to an Ohmic ($s=1$) bath are investigated. In Fig.~\ref{fig4}(c) and (f), $\Delta/\omega$ and $\alpha$ are fixed, and $s$ is varied to examine the dynamic differences caused by  sub-Ohmic (panel (c)) and super-Ohmic (panel (d)) baths. All populations in Fig.~\ref{fig4} are evaluated for $v/\omega^2$ = 1, $\omega_c/\omega = 10$ with the wave function initialized in $|{+, {\rm vac}}\rangle$.

In Figs.~\ref{fig4} (a) and (d), we fix $\alpha$ at 0.002, and change the tunneling strength $\Delta/\omega$ from 0 to 0.5. The population dynamics in  Fig.~\ref{fig4}(a)  can be separated into two phases. During the first stage,  ${\mathcal P}_{-}(t)$  exhibits a rise near $\omega t = 0$  and is mainly affected by the tunneling: the higher $\Delta/\omega$, the larger ${\mathcal P}_{-}(t)$. Comparing the curves with $\Delta/\omega = 0$ and $\Delta/\omega = 0.1$, for example, we see that the difference between them is small (cf. the discussion of Fig.~\ref{fig3}(e)). When $\omega t > 0$, the changes in transition probabilities are mostly due to the coupling to the bath. This phase is characterized by an overall increase of ${\mathcal P}_{-}(t)$ which is superimposed with coherent St\"uckelberg oscillations of decreasing amplitude and period (cf. Ref. \cite{Vitanov99}). Note that the gradient (that is, the rate of increase) of  transition probabilities is decided by the spectral density function. Since $\alpha$ is fixed in Fig.~\ref{fig4}(a), the ${\mathcal P}_{-}(t)$ curves for different $\Delta/\omega$ are roughly parallel to each other  at $\omega t > 0$.
The situation is different for ${\mathcal P}_{0}(t)$  depicted in Fig.~\ref{fig4}(d). If $\Delta/\omega$ is small, the coupling to the bath enhances ${\mathcal P}_{0}(t)$ at $\omega t > 0$. This is similar to the behaviour of ${\mathcal P}_{-}(t)$. As $\Delta/\omega$ becomes larger,  ${\mathcal P}_{0}(t)$ starts decreasing at $\omega t > 0$. This indicates a strong dependence of the bath-induced dissipation on  $\Delta/\omega$. Such a combined dissipation + tunneling affect is absent if  the 3L-BTM is  coupled to a single harmonic mode.  In this latter case, if the wave function is initialized in $|{+, {\rm vac}}\rangle$, the LZ transition in  ${\mathcal P}_{0}(t)$ induced by the tunneling  $\Delta/\omega$ is independent of the LZ transition induced by $\Lambda/\omega$, and the steady state transition probability is simply a sum of the two probabilities (see Fig.~\ref{fig3}). The mutual ``entanglement" of the tunneling ($\Delta$) and bath ($\alpha$) induced effects is also not seen in dissipative two-level LZ models \cite{LZ_HO1}. This makes the bath-tunneling entanglement a signature of the LZ transitions in dissipative  3L-BTMs.

In Fig.~\ref{fig4}(b) and (e), $\Delta$ is fixed at 0.1, and the bath coupling strength $\alpha$ is changed from 0.002 to 0.01. In Fig.~\ref{fig4}(b), ${\mathcal P}_{-}(t)$ is plotted with respect to time. Changing of $\alpha$ causes  variation of the gradient of ${\mathcal P}_{-}(t)$ between $\omega t = 0$ and $\omega t = 20$. As $\alpha$ becomes larger, ${\mathcal P}_{-}(t)$ increases more rapidly between $\omega t = 0$ and $\omega t = 20$. This leads to higher values of ${\mathcal P}_{-}(t)$ in the  steady state.  Fig.~\ref{fig4}(e) displays the ${\mathcal P}_{0}(t)$ evolution. Comparing ${\mathcal P}_{-}(t)$ (Fig.~\ref{fig4}(b)) and ${\mathcal P}_{0}(t)$ (Fig.~\ref{fig4}(e)), we arrive at the following interesting observation.  When $\alpha$ is small, ${\mathcal P}_{0}(t)$ is larger than ${\mathcal P}_{-}(t)$  throughout the entire time evolution. If $\alpha$ becomes larger, ${\mathcal P}_{-}(t)$ increases, too, but ${\mathcal P}_{0}(t)$ decreases. When $\alpha = 0.1$, ${\mathcal P}_{-}(\infty) \approx 1$, while ${\mathcal P}_{0}(\infty) \approx 0$. This again indicates that the indirect $|{+}\rangle \rightarrow |{-}\rangle$ transition is dominant when the coupling strength is large.

In Figs.~\ref{fig4}(c) and (f), $\alpha$ and $\Delta$ are fixed at 0.002 and 0.1, and $s$ is changed from 0.5 to 1.5. Fig.~\ref{fig4}(c) corresponds to sub-Ohmic baths with  $s \le 1$, while  Fig.~\ref{fig4}(f) corresponds to  super-Ohmic baths with $s \ge 1$. In both sub- and super-Ohmic regimes, increasing $s$ leads to higher values of  ${\mathcal P}_{-}(\infty)$. Interestingly, equal increments in increasing $s$ yield lower ${\mathcal P}_{-}(\infty)$ in the sub-Ohmic regime than in the super-Ohmic regime. The reason is that the value of ${\mathcal P}_{-}(\infty)$ is governed by the integral of the spectral density function $J(\omega)$ over the entire range of frequencies. With equal increments in increasing $s$, the change of this integral is smaller in the sub-Ohmic regime in comparison with the super-Ohmic regime. However, as $s$ becomes smaller, the situation changes and sub-Ohmic baths cause faster increase of ${\mathcal P}_{-}(t)$, i.e., produce larger gradients of ${\mathcal P}_{-}(t)$  between $\omega t = 0$ and $\omega t = 20$. This is not observed in the super-Ohmic regime, where the gradient of ${\mathcal P}_{-}(t)$ between $\omega t = 0$ and $\omega t = 20$ is approximately the same for all $s$. The explanation is similar. In the sub-Ohmic regime, as $s$ becomes smaller, the $J(\omega)$ maximum shifts towards lower frequencies, which causes faster increase in ${\mathcal P}_{-}(t)$.

\section{Conclusion}

By employing the multi-D${_2}$ Davydov Ansatz, we  performed numerically ``exact'' simulations of the dissipative 3L-BTM dynamics. We  considered a bare 3L-BTM  coupled to a single harmonic mode as well as a bare 3L-BTM  coupled to a boson bath with Ohmic, sub-Ohmic and super-Ohmic spectral densities. With the aid of the energy diagrams, we  developed a useful qualitative method of characterizing and interpreting population-transfer pathways in dissipative 3L-BTMs. This method has revealed mechanisms behind various LZ transitions in the 3L-BTM and uncovered their contributions to the steady-state populations.

We have shown that vibrational splittings of the electronic levels of the 3L-BTM cause nontrivial crossing patterns in the energy diagram which can be understood by inspecting sequential wavepacket scatterings on the relevant electronic/vibrational states. These scattering processes can be  directly linked to the steady-state populations. We have demonstrated that the 3L-BTM dynamics is very sensitive to tunneling  strengths, system-bath couplings, and characteristic frequencies of the bath. In particular, the presence of boson states breaks the SU2 symmetry of the bare 3L-BTM, causing asymmetry of the pathways leading from the initial $|0\rangle$ state to the final $|{+}\rangle$ and $|{-}\rangle$ states. In general, we found profound differences between the time evolution of the original two-level LZ model and the present 3L-BTM. In certain cases, however, the dynamics initiated in the lowest state  $|{-}\rangle$ of the 3L-BTM is almost insensitive to the presence of the upper  $|{+}\rangle$ state, which renders the 3L-BTM to behave like an effective two-level LZ model.

Our simulations prove that sub-Ohmic, Ohmic and super-Ohmic boson baths have different impact on the 3L-BTM dynamics. In particular, rise times, local maxima and subsequent decays of the 3L-BTM  populations depend significantly on the parameters $\alpha$ and $s$ specifying the bath spectral density. Hence the phenomenological Lindblad-like descriptions reducing all multifaceted bath-induced phenomena to a single relaxation rate are rendered inadequate for reproducing actual dynamics of dissipative 3L-BTMs. The numerically accurate methodology with the multi-D$_2$ Davydov Ansatz developed in this work can help to interpret experiments on spin-1 systems, facilitate the development of QED devices based on these systems, and provide the guidance for engineering and optimizing these devices.

Note, finally, that the computational efficiency of the multi-D${_2}$ Ansatz does not crucially depend on specific values of the system and bath parameters. By adopting the Thermo Field Dynamics framework, the multi-D${_2}$ Ansatz can be turned into accurate  simulator of LZ systems at finite temperatures \cite{TFD1, TFD2}. Hence the versatile multi-D${_2}$ machinery can become a method of choice for simulations of general multilevel dissipative LZ systems.

\section*{Acknowledgments}
The authors thank Lu Wang, Kewei Sun, Fulu Zheng, and Frank Grossmann for useful discussion, and Zongfa Zhang for providing access to computational resources.
Support from Nanyang Technological University ``URECA" Undergraduate Research Programme and
the Singapore Ministry of Education Academic Research Fund Tier 1 (Grant No.~RG87/20) is gratefully acknowledged.

\section*{Author Declarations}

\subsection*{Conflict of Interest}
The authors have no conflicts to disclose.

\section*{Data Availability}
The data that support the findings of this study are available from the corresponding author upon reasonable request.

\appendix

\section{Convergence proof}\label{Appendix Z}

Here we demonstrate the numerical convergence of the calculations of the present work. The chosen values of multiplicities of the multi-D$_2$ Ansatz and of other parameters ensure that the results presented in Sec. \ref{RD} are converged.

\subsection{Single-mode 3L-BTM}

\begin{figure}[ht]
\centerline{\includegraphics[width=85mm]{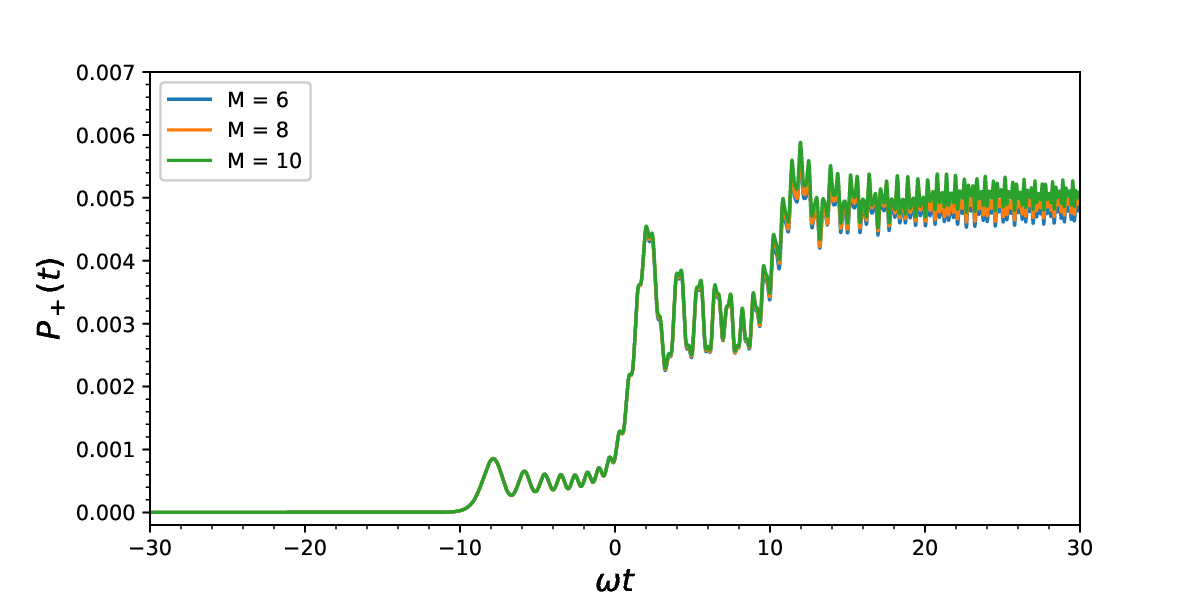}}
\caption{The time evolution of  ${\mathcal P}_{+}(t)$ from $\omega t = -30$ to $\omega t = 30$ for different multiplicities $M$. The calculation is initialized in the state  $|{-}\rangle$. The remaining parameters are as follows: $\Delta/\omega = 0.1$, $v/\omega^2 = 1$, $\Lambda/\omega = 0.1$, and $\Omega/\omega = 10$.}
\label{figs1}
\end{figure}

In Sec. \ref{Om}, we consider the 3L-BTM coupled to a single harmonic mode. The convergence of the results is determined by the multiplicity $M$ of the multi-D$_2$ Ansatz. When $M$ is large enough, the results are independent of $M$ and convergence is reached. Fig.~\ref{figs1} depicts population ${\mathcal P}_{+}(t)$ calculated for different $M$ . The remaining parameters are fixed: $\Delta/\omega = 0.1$, $v/\omega^2 = 1$, $\Lambda/\omega = 0.1$, and $\Omega/\omega = 10$.It can be seen that difference between ${\mathcal P}_{+}(t)$ curves for $M=8$  and 10 is negligible. 

\subsection{3L-BTM coupled to harmonic bath}

\begin{figure}[ht]
\centerline{\includegraphics[width=85mm]{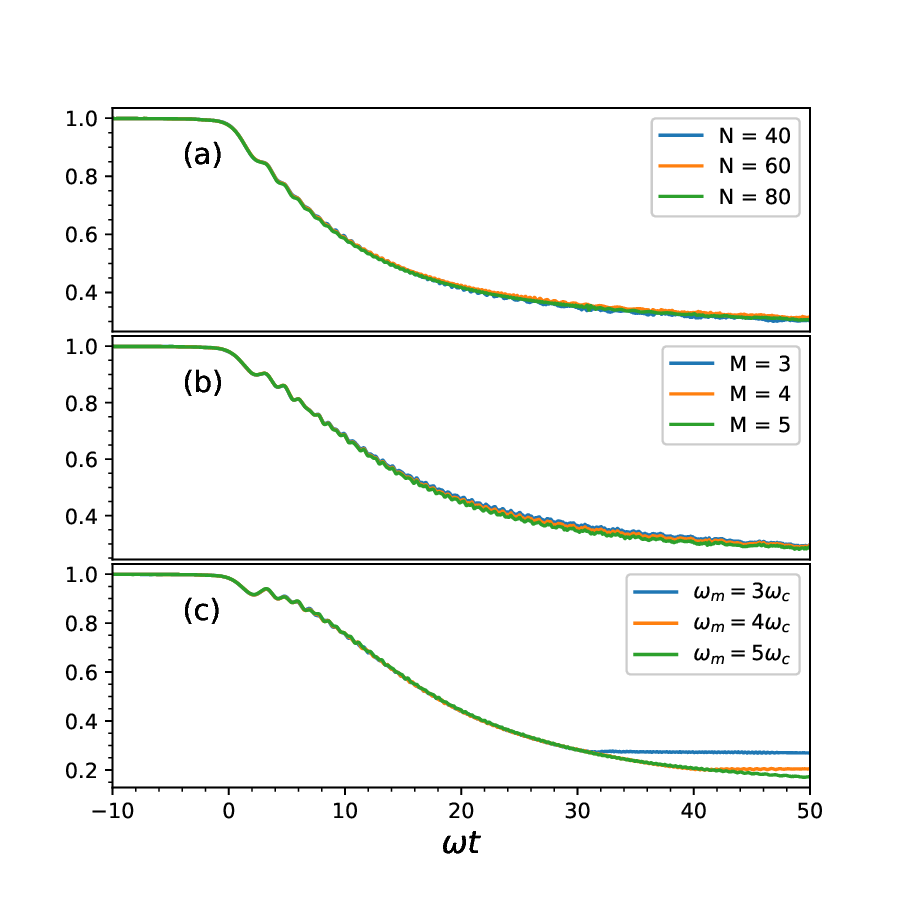}}
\caption{The convergence test to $N$, $M$ and $\omega_m/\omega$ of the time evolution of ${\mathcal P}_{+}$ from $\omega t = -10$ to $\omega t = 50$. In each one of the subplots, only one of the parameters is changed, the rest of the parameters are fixed. (a). $N = $ 40, 60 and 80. $M$ = 4, $\omega_m = 5\omega_c$ and $s$ = 0.5. (b). $M$ = 3, 4 and 5. $N$ = 40, $\omega_m = 5\omega_c$ and $s$ = 1. (c).  $\omega_m = 3\omega_c, 4\omega_c, 5\omega_c$. $M$ = 4, $N$ = 40 and $s$ = 1.75. The rest of the parameters are consistent for all of the subplots: $\Delta = 0.1$, $v = 1$, $\alpha = 0.002$ and $\omega_c = 10\omega$.}
\label{figs2}
\end{figure}

In Sec. \ref{Diss}, we consider the 3L-BTM coupled to a harmonic bath. In this case, the results depend on the multi-D$_2$ multiplicity $M$ as well as on the parameters specifying discretization of the bath spectral density, viz. the maximum frequency $\omega_m/\omega$ and the discrete mode number $N$. It is known that the smaller is the exponent $s$ in the bath spectral density of Eq. (\ref{EQ6}), the higher $N$  is required to reach the convergence. Therefore in Fig.~\ref{figs2}(a) we choose the smallest $s$ that is used in the simulations of Sec. \ref{Diss},  $s = 0.5$. It can be seen that different choices of $N$ have no significant effect on ${\mathcal P}_{+}(t)$. In Fig.~\ref{figs2}(b), a similar procedure is performed for $M$. It can be seen that $M = 3$ is already sufficient to yield the converged results. In Fig.~\ref{figs2}(c), shows ${\mathcal P}_{+}(t)$ for different $\omega_m$. It is observed that  ${\mathcal P}_{+}(t)$ stops changing if $\omega t < \omega_m/\omega$. In other words, the choice of $\omega_m/\omega$ has no influence on the dynamics if  $\omega t < \omega_m/\omega$. The indicates that if  $\omega_m/\omega > \omega t_{\rm max}$ where $t_{\rm max}$ is the final time of the calculation, the results are converged.

\subsection{Comparison of different discretization methods}

\begin{figure}[ht]
\centerline{\includegraphics[width=85mm]{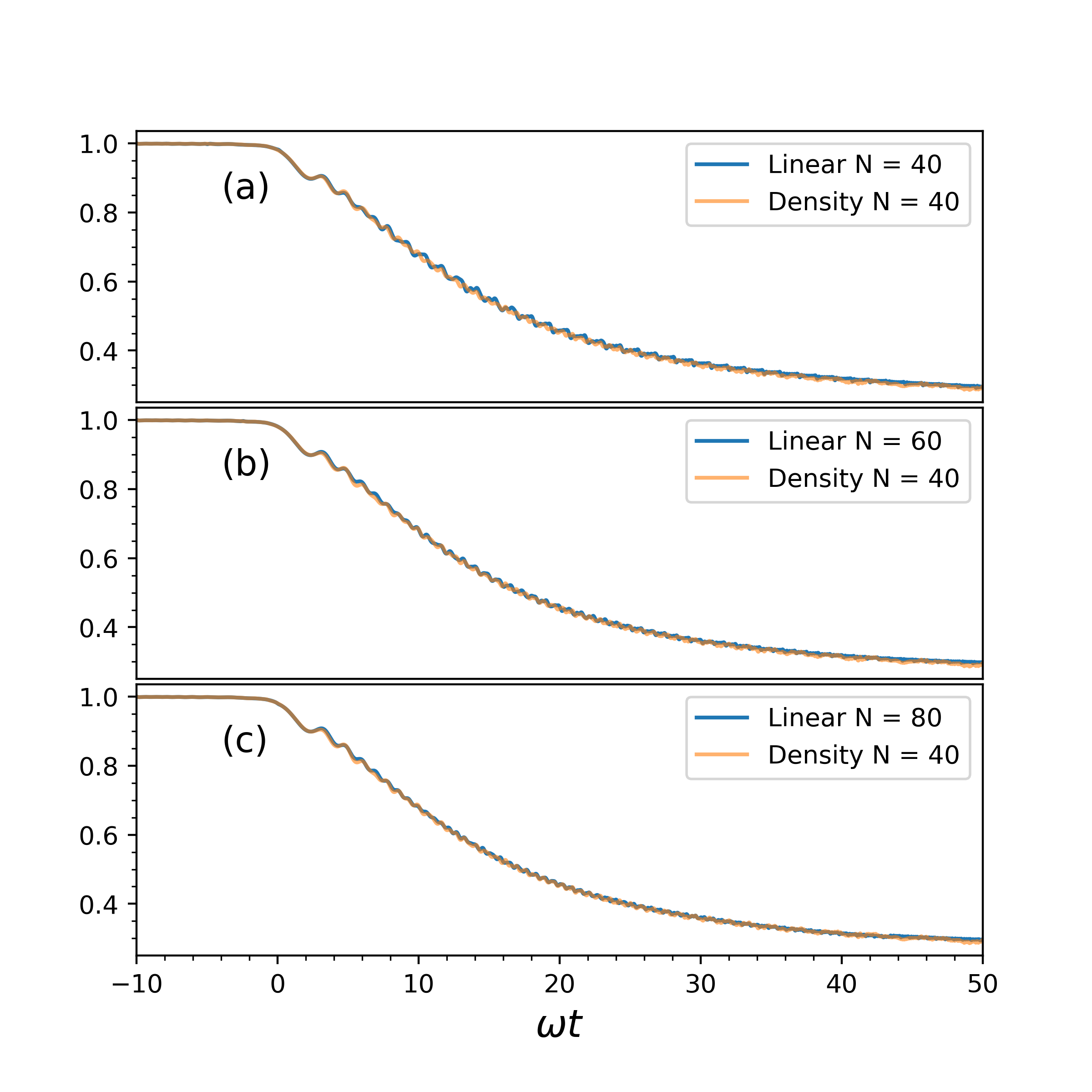}}
\caption{A comparison between density discretization and the linear discretization by testing the time evolution of  ${\mathcal P}_{+}$ from $t = -10 \omega^{-1}$ to $t = 50 \omega^{-1}$ with different discrete mode number $N$. (a). 40 discrete modes are used for both discretization methods. (b). 40 modes are used for density discretization, and 60 modes are used for linear discretization. (b). 40 modes are used for density discretization, and 80 modes are used for linear discretization. The rest of the parameters are consistent for all of the subplots: $\Delta = 0.1$, $v = 1$, $s = 1$, $\alpha = 0.002$ and $\omega_c = 10$.}
\label{figs3}
\end{figure}

In order to validate the performance of the density discretization method introduced in Sec. \ref{Discr}, we benchmark it against the commonly used linear discretization method for the 3L-BTM coupled to the Ohmic bath.
Fig.~\ref{figs3}(a) shows ${\mathcal P}_{+}(t)$ calculated by both methods by using $N=40$ discrete modes.
The density discretization method yields correct smooth evolution of ${\mathcal P}_{+}(t)$ (see below),  while spurious stair-like pattern caused by undersampling can be seen in ${\mathcal P}_{+}(t)$  calculated by the linear discretization method. As the number of modes in the linear discretization method increases (Fig.~\ref{figs3}(b)), the spurious structures smoothen out. Finally, ${\mathcal P}_{+}(t)$ produced by the linear discretization method with 80 modes overlaps with the ${\mathcal P}_{+}$ calculated by the density method with 40 discrete modes (Fig.~\ref{figs3}(c)). This proves that the density discretization method  converges faster than the linear discretization method.

\onecolumngrid
\section{Equations of motions for the multi-D$_2$ Ansatz}\label{Appendix A}

For $A_{m+}^{\ast}$:
\begin{eqnarray}
&&i\sum_{n=1}^{M} \Big[\dot{A}_{n+}+ A_{n+}\sum_{k} (\dot{\alpha}_{nk} {\alpha}^{\ast}_{m{k}}-\frac{1}{2}\dot{\alpha}_{nk}\alpha_{nk}^{\ast}-\frac{1}{2}{\alpha}_{nk}\dot{\alpha}_{nk}^{\ast})\Big]S_{mn}\nonumber \\
&&= \sum_{n=1}^{M}\Big[A_{n+}vt + A_{n0}\Big({\Delta}+\sum_k \eta_{k}({\alpha}^{\ast}_{mk}+{\alpha}_{nk})\Big)+A_{n+}\Big(\sum_k\omega_k\big({\alpha}^{\ast}_{mk}{\alpha}_{nk})\Big)\Big]S_{mn}
\end{eqnarray}

For $A_{m0}^{\ast}$:
\begin{eqnarray}
&&i\sum_{n=1}^{M} \Big[\dot{A}_{n0}+ A_{n0}\sum_{k} (\dot{\alpha}_{nk} {\alpha}^{\ast}_{m{k}}-\frac{1}{2}\dot{\alpha}_{nk}\alpha_{nk}^{\ast}-\frac{1}{2}{\alpha}_{nk}\dot{\alpha}_{nk}^{\ast})\Big]S_{mn}\nonumber \\
&&= \sum_{n=1}^{M}\Big[(A_{n+}+A_{n-})\Big({\Delta}+\sum_k \eta_{k}({\alpha}^{\ast}_{mk}+{\alpha}_{nk})\Big)+A_{n0}\Big(\sum_k\omega_k\big({\alpha}^{\ast}_{mk}{\alpha}_{nk})\Big)\Big]S_{mn}
\end{eqnarray}

For $A_{m-}^{\ast}$:
\begin{eqnarray}
&&i\sum_{n=1}^{M} \Big[\dot{A}_{n-}+ A_{n-}\sum_{k} (\dot{\alpha}_{nk} {\alpha}^{\ast}_{m{k}}-\frac{1}{2}\dot{\alpha}_{nk}\alpha_{nk}^{\ast}-\frac{1}{2}{\alpha}_{nk}\dot{\alpha}_{nk}^{\ast})\Big]S_{mn}\nonumber \\
&&= \sum_{n=1}^{M}\Big[-A_{n-}vt + A_{n0}\Big({\Delta}+\sum_k \eta_{k}({\alpha}^{\ast}_{mk}+{\alpha}_{nk})\Big)+A_{n0}\Big(\sum_k\omega_k\big({\alpha}^{\ast}_{mk}{\alpha}_{nk})\Big)\Big]S_{mn}
\end{eqnarray}

For  $\alpha^{\ast}_{m{k}}$:
\begin{eqnarray}
&&i\sum_{m,n}^{M} \sum_s^{+, -, 0}\Big[ A_{ms}^{\ast}\dot{A}_{ns}+  A_{ms}^{\ast}{A}_{ns} \Big(  \dot{\alpha}_{nk}+\alpha_{nk}\sum_{k^\prime} \dot{\alpha}_{nk^\prime} {\alpha}^{\ast}_{mk^\prime}-\frac{1}{2}\sum_{k^\prime}(\dot{\alpha}_{nk^\prime}\alpha_{nk^\prime}^{\ast}+{\alpha}_{nk^\prime}\dot{\alpha}_{nk^\prime}^{\ast})\Big)\Big]S_{mn} \nonumber \\
&&= \sum_{m,n}^{M}\Big[\Big( A_{m+}^{\ast}A_{n+}-A_{m-}^{\ast}A_{n-}\Big){vt}\alpha_{nk}+\sum_s^{+, -, 0} A_{ms}^{\ast}A_{ns}\Big(\omega_{k}{\alpha}_{nk}+\alpha_{nk}\sum_{k^{\prime}}\omega_{k^{\prime}}{\alpha}^{\ast}_{m{k^{\prime}}}{\alpha}_{n{k^{\prime}}}\Big)\nonumber \\
&&+\Big( A_{m0}^{\ast}A_{n+}+A_{m+}^{\ast}A_{n0}+A_{m-}^{\ast}A_{n0}+A_{m0}^{\ast}A_{n-}\Big)\Big({\Delta}\alpha_{nk}+ \eta_{k}+\alpha_{nk}\sum_{{k^{\prime}}}\eta_{{k^{\prime}}}({\alpha}^{\ast}_{m{k^{\prime}}}+{\alpha}_{n{k^{\prime}}})\Big)\Big]S_{mn}\nonumber \\
\end{eqnarray}

\twocolumngrid

\end{document}